\documentclass[longbibliography, reprint, aps, nofootinbib, superscriptaddress, floatfix]{revtex4-2} 
\usepackage{amsmath, amssymb, mathtools, graphicx, dcolumn,  microtype}
\usepackage[noautoscale]{youngtab}\Yboxdim 6.5pt\Ylinethick 0.5pt  
\usepackage[bb=dsfontserif]{mathalpha}   
\usepackage[dvipsnames]{xcolor}\definecolor{darkgreen}{rgb}{0,0.4,0.2}  
\usepackage{tikz} 
\usetikzlibrary{arrows.meta}   
\usepackage{orcidlink} 
\usepackage{hyperref}\hypersetup{colorlinks=true, urlcolor=Blue, citecolor=Blue, linkcolor=Blue}
\graphicspath{{Figures/}}
 
\newcommand{\T}{\mathbb{T}} 
\newcommand{\Z}{\mathbb{Z}} 
\newcommand{\K}{\ensuremath{\mathbb{K}}}  

\newcommand{\U}{\textrm{U}}
\newcommand{\SU}{\textrm{SU}}
\newcommand{\USp}{\textrm{USp}}

\begin{document}

\title{Three family supersymmetric Pati-Salam model \\
from rigid intersecting D6-branes}

\author{Adeel Mansha\,\orcidlink{0000-0002-1183-0355}}
\email{adeelmansha@alumni.itp.ac.cn}
\affiliation{College of Physics and Optoelectronic Engineering, Shenzhen University, Shenzhen 518060, P.R. China}

\author{Mudassar Sabir\,\orcidlink{0000-0002-8551-2608}}
\email{mudassar.sabir@uestc.edu.cn} 
\affiliation{School of Physics, University of Electronic Science and Technology of China, Sichuan 611731, Chengdu, P.R. China}

\author{Tianjun Li\,\orcidlink{0000-0003-1583-5935}}
\email{tli@itp.ac.cn}
\affiliation{CAS Key Laboratory of Theoretical Physics, Institute of Theoretical Physics, Chinese Academy of Sciences, Beijing 100190, P.R. China}
\affiliation{School of Physical Sciences, University of Chinese Academy of Sciences, Beijing, P.R. China}

\author{Luyang Wang\,\orcidlink{0000-0001-9400-7331}}
\email{wangly@szu.edu.cn}
\affiliation{College of Physics and Optoelectronic Engineering, Shenzhen University, Shenzhen 518060, P.R. China}

\begin{abstract}
We construct, for the first time, a three-family $\mathcal{N}=1$ supersymmetric Pati-Salam model from rigid intersecting D6-branes on a factorizable $\mathbb{T}^6/(\mathbb{Z}_2\times \mathbb{Z}_2')$ orientifold with discrete torsion. The factorizable geometry allows for explicit control over rigid cycles and moduli stabilization. We can break the Pati-Salam gauge symmetry down to the Standard Model (SM) gauge symmetry via the supersymmetry preserving Higgs mechanism, generate the SM fermion masses and mixings, and break the supersymmetry via gaugino condensations in the hidden sector.
\end{abstract}

\maketitle
 
\noindent\textbf{Introduction.}
A great challenge in string phenomenology has been to construct the realistic string vacua, which can give the low energy supersymmetric Standard Models (SMs) without exotic particles, and stabilize the moduli fields.  We can connect such string models to the low energy realistic particle physics via renormalization group equation running, and thus probe these models at the Large Hadron Collider (LHC) and the future Colliders, etc. We would like to emphasize that the semi-realistic supersymmetric Standard-like models and Grand Unified Theories (GUTs) have been constructed extensively in Type IIA theory on the $\T^6/(\Z_2\times \Z_2)$ orientifold~\cite{Cvetic:2001tj, Cvetic:2001nr, Cvetic:2002pj, Cvetic:2004ui, Blumenhagen:2005mu, Chen:2005mj, Chen:2006gd, Chen:2006ip, Chen:2007px, Chen:2007zu, He:2021kbj, He:2021gug, Li:2022cqk, Sabir:2022hko, Li:2019nvi, Li:2021pxo, Mansha:2022pnd, Mansha:2023kwq, Sabir:2024mfv, Sabir:2024cgt, Sabir:2024jsx, Mansha:2024yqz}.  

The key point is that only Pati-Salam like models can explain all the Yukawa couplings~\cite{Cvetic:2004ui}. Such kind of models can also be constructed in the heterotic string framework~\cite{Assel:2010wj, Faraggi:2020wld}. In heterotic models, the gauge group arises from the breaking of $\text{E}_8 \times \text{E}_8$ via internal bundles and orbifold twists, with matter spectra determined by the bundle structure. And thus there exists gauge coupling unification naturally. In contrast, the intersecting D6-brane models generate gauge groups from stacks of branes wrapping factorizable three-cycles, with chiral matter localized at brane intersections. This approach provides a clear geometric understanding of gauge symmetries, chirality, and coupling hierarchies, and allows for the systematic construction of rigid cycles without adjoint matter. Furthermore, couplings such as $\mu$-terms and non-perturbative neutrino masses can arise from D-brane instantons. Global consistency conditions such as RR tadpole and K-theory constraints are manifest, ensuring tight control over the resulting low-energy spectrum.

The three family supersymmetric Pati-Salam models have been constructed systematically first in~\cite{Cvetic:2004ui}. The SM fermion masses and mixings in some models have been explained explicitly~\cite{Chen:2007px, Chen:2007zu, Sabir:2022hko, Sabir:2024cgt}. In particular, a systematic method was proposed to construct all the three-family $\mathcal{N}=1$ supersymmetric Pati-Salam models where the Pati-Salam gauge symmetry can be broken down to the SM gauge symmetry via the D-brane splitting and supersymmetry preserving Higgs mechanism~\cite{He:2021kbj, He:2021gug}. However, in these semi-realistic models, there exist three adjoint multiplets for $\U(N)$ gauge symmetries since the D6-branes are not rigid. And the models from the rigid intersecting D6-branes are still very far from realistic~\cite{Blumenhagen:2005tn, Forste:2008ex, Blumenhagen:2002wn, Ibanez:2008my, Blumenhagen:2006xt, Ibanez:2006da, Abel:2006yk, Blumenhagen:2007zk}. Therefore, how to construct the three-family $\mathcal{N}=1$ supersymmetric Pati-Salam models from rigid intersecting D6-branes is a big challenge. 

In Ref.~\cite{Blumenhagen:2005tn} it was first shown that the global $\mathcal{N} = 1$ chiral models can be constructed from the rigid intersecting D6-branes on a factorizable $\T^6/(\Z_2\times \Z_2')$ orientifold with discrete torsion. However, the consistent three-generation models in this setup with factorizable tori were very difficult to engineer, and the examples in \cite{Blumenhagen:2005tn} generically had four generations. In the case of non-rigid branes, we need at least one torus to be tilted to generate three-generation models on the mirror orbifold $\T^6/(\Z_2\times \Z_2)$~\cite{Cvetic:2001tj, Cvetic:2001nr, Cvetic:2002pj, Cvetic:2004ui, Blumenhagen:2005mu, Chen:2005mj, Chen:2006gd, Chen:2006ip, Chen:2007px, Chen:2007zu, He:2021kbj, He:2021gug, Li:2022cqk, Sabir:2022hko, Li:2019nvi, Li:2021pxo, Mansha:2022pnd, Mansha:2023kwq, Sabir:2024cgt}. However, in the present setup with rigid 3-cycles, no consistent three-generation models employing tilted-tori are known. In Ref.~\cite{Forste:2008ex} a first example of a three-generation model was indeed presented, however it was constructed on non-factorizable tori. The equivalent model with factorizable tori does not satisfy $\mathcal{N}=1$ supersymmetry conditions. 

The main advantage of using factorizable toroidal orbifolds is the enhanced control over D6-brane configurations and the systematic construction of rigid three-cycles. On factorizable spaces, D-brane cycles decompose into products of 1-cycles on each two-torus, simplifying the computation of intersection numbers, tadpole conditions, and chiral spectra. In contrast, non-factorizable geometries often obscure the localization of exceptional cycles and complicate the classification of rigid three-cycles, as discussed in~\cite{Forste:2007zb}. Furthermore, the intersection number formulas and tadpole cancellation conditions depend on the choice of lattice (e.g., A- or B-type), which is more transparent in factorizable settings. In this letter, we for the first time construct a three-family $\mathcal{N}=1$ supersymmetric Pati-Salam model with factorizable tori from rigid intersecting D6-branes on a factorizable $\T^6/(\Z_2\times \Z_2')$ orientifold utilizing only the rectangular tori (A-type lattice), and briefly study its phenomenological consequences. \\

\noindent\textbf{Model Building from Rigid Intersecting D6-Branes.}
Discrete torsion leads to rigid D6-branes wrapping fractional, and exceptional three-cycle collapsed to a lower dimensional fixed cycle in the orbifold limit of the Calabi-Yau space~\cite{Blumenhagen:2002wn}. As the D6-branes are rigid, adjoint moduli disappear from the open string spectra: the irregularity of gauge symmetry breaking is removed along the flat direction, and one-loop beta functions are improved. Rigid branes allow instanton contributions to the K{\"a}hler potential, and superpotential generating $\mu$-terms~\cite{Ibanez:2008my}, non-perturbative neutrino masses~\cite{Blumenhagen:2006xt,Ibanez:2006da}, forbidden Yukawa couplings~\cite{Abel:2006yk,Blumenhagen:2007zk} or might even trigger supersymmetry breaking~\cite{Cvetic:2008mh}. 

\begin{figure}[!t]
\centering
\begin{tikzpicture}[x=3cm, y=3cm, scale=2/3] 
		\begin{scope}
			\draw[-Latex] (0,0) -- (1.5,0) node[right] {$x^I$};
			\draw[-Latex] (0,0) -- (0,1.5) node[above] {$y^I$};
			\draw[dotted] (0,.5) -- (1,.5) ;
						  
			\draw (0,0) rectangle (1,1);
			\draw[thick, Red] (0,0) -- (1,0) node[midway, below] {$[a^I]$}; 
			\draw[thick, RoyalBlue] (0,0) -- (0,1) node[midway, left] {$[b^I]$}; 
						   
			\node at (.5,1.5) {$\beta^I = 0$}; 
			\node at (1,0) [below]{$R_1^I$};
			\node at (0,1) [left]{$R_2^I$};
			\fill[darkgreen] (0,0) circle (2pt) node[above right]{1};
			\fill[darkgreen] (0,.5) circle (2pt) node[above right]{2};
			\fill[darkgreen] (.5,0) circle (2pt) node[above]{3};
			\fill[darkgreen] (.5,.5) circle (2pt) node[above]{4};
		\end{scope}
		\begin{scope}[xshift=6.7cm]
			\draw[-Latex] (0,0) -- (1.5,0) node[right] {$x^I$};
			\draw[-Latex] (0,0) -- (0,1.5) node[above] {$y^I$};
			\draw[dotted] (0,.5) -- (1,.5) ;
			\draw[dotted] (0,1) -- (1,1) ;
						
			\draw (0,0) -- (1,.5) -- (1,1.5) -- (0,1);
			\draw[thick, Red] (0,0) -- (1,0) node[midway, below] {$[a^I]$}; 
			\draw[thick, RoyalBlue] (0,0) -- (0,1) node[midway, left] {$[b^I]$}; 
			\draw[thick, Green] (0,0) -- (1,.5) node[near end, below] {$[a'^I]$};
						  
			\node at (.5,1.5) {$\beta^I = 1$}; 
			\draw (1,-0.025) -- (1,0.025) ;
			\node at (1,0) [below]{$R_1^I$}; 
			\node at (0,1) [left]{$R_2^I$}; 
			\fill[darkgreen] (0,0) circle (2pt) node[above right, yshift=0.05cm]{1};
			\fill[darkgreen] (0,.5) circle (2pt) node[above right]{2};
			\fill[darkgreen] (.5,.25) circle (2pt) node[above]{3};
			\fill[darkgreen] (.5,.75) circle (2pt) node[above]{4};
		\end{scope}
	\end{tikzpicture}  
\caption{The $\Z_2$ invariant {\bf a}-type (left) and {\bf b}-type (right) lattices.
$\Z_2$ fixed points are depicted in dots.
The $\mathcal{R}$ invariant $x^I$ axis is along the 1-cycle $[a^I] - \frac{\beta^I}{2} [b^I]$ with $\beta^I=0$ for the {\bf a}-type lattice and $\beta^I=1$ for the {\bf b}-type lattice.
$\mathcal{R}$ acts as reflection along the $y^I$ axis, which is spanned by the 1-cycle $[b^I]$.
For the {\bf a}-type lattice, all $\Z_2$ fixed points are invariant under $\mathcal{R}$, whereas
for the {\bf b}-type lattice, only 1 and 2 are invariant while 3 and 4 exchange as $3 \stackrel{\mathcal{R}}{\leftrightarrow} 4$.}
\label{fig:Z2-lattice}
\end{figure}
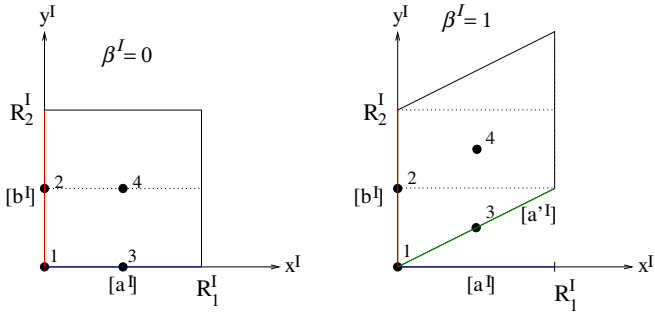
 
\begin{table}[!t]
\renewcommand{\arraystretch}{1.4}
\centering
\begin{tabular}{|c|c|}
\hline {\bf Sector} & {\bf Representation} \\
\hline\hline
$aa$   & $\text{U}(N_a/2)$ vector multiplet  \\ 
\hline $ab+ba$   & $ I_{ab}\,(\yng(1)_{a},\overline{\yng(1)}_{b}) = \Pi_a\circ \Pi_b $ \\
\hline $ab'+b'a$ & $ I_{ab'}\,(\yng(1)_{a},\yng(1)_{b})= \Pi_a \circ \Pi_b' $ \\
\hline $aa'+a'a$ & $ n_{\yng(2)}= \frac{1}{2} (\Pi_a \circ \Pi_a' - \Pi_{a} \circ \Pi_{O6}) $  \\
                 & $ n_{\yng(1,1)_{}}= \frac{1}{2} (\Pi_a \circ \Pi_a' + \Pi_{a} \circ \Pi_{O6}) $  \\
\hline
\end{tabular}
\caption{Chiral spectrum for intersecting D6-branes at angles. Positive intersection numbers refer to the left-handed fermions.}
\label{tab:spectrum}
\end{table}

In setup~\cite{Blumenhagen:2005tn}, the six-torus is factorizable as $\T^6=\T^2\times \T^2 \times \T^2$, and the orbifold group is $\mathbb{Z}_2\times\mathbb{Z}'_2$, where each $\mathbb{Z}_2$-factor inverts two two-tori. In addition to the untwisted cycles, there are 48 independent collapsed three-cycles from the twisted sectors. Thus, in this setup, there are, in total, 104 3-cycles, i.e., $b_3(\T^6/(\Z_2\times \Z_2'))=2 + 2h^{21}_{unt}+ 2h^{21}_{tw}$, with $(h^{21}_{unt}, h^{21}_{tw})=(3, 48)$. For each sector, we denote the 16 fixed points on $(\T^2_1 \times \T^2_2)/\Z_2$ by $[e^g_{i,j}]$, with $g \in \{\Theta, \Theta', \Theta \Theta'\}$, and  $i,j\in\{1,2,3,4\}$ as depicted in Fig.~\ref{fig:Z2-lattice}. After blowing up the orbifold singularities, these become two-cycles with the topology of $\mathbb{S}^2$. Each such four-dimensional $\T^4/\Z_2$ is an orbifold of \K3 before taking the other elements of the orientifold group into account. There are three $\mathbb{Z}_2$-twisted sectors with sixteen fixed points each.  The three two-tori have radii $R_{1,2}^{(i)}$ along the $x^i, y^i$-axes, $i\in\{1,2,3\}$. The tori may ($\beta^i=1$)\footnote{Unlike \cite{Blumenhagen:2005tn}, we take $\beta=1$ and not 1/2 for the tilted torus.} or may not ($\beta^i=0$) be tilted. There are four orbifold fixed points on each torus, at $(0,0)$, $(0,R_2^{(i)}/2)$,  $(R_1^{(i)}/2,\beta^i R_2^{(i)}/4)$ and $(R_1^{(i)}/2,(2+\beta^i) R_2^{(i)}/4)$. They will be labelled fixed points $1,2,3$ and $4$. All this geometrical data is shown in Fig. \ref{fig:Z2-lattice} for an untilted and a tilted torus. The orientifold group is $\Omega \mathcal{R} (-1)^{F_L}$, where $F_L$ is the left-moving spacetime fermion number, $\Omega$ is world-sheet parity, and $\mathcal{R}$ is an isometric antiholomorphic involution of internal space acting on the K{\"a}hler class $J$ and three-from $\Omega_3$ as
\begin{equation}
    \mathcal{R} J = -J,\ \ \mathcal{R}\Omega_3 = e^{2i\phi}\bar{\Omega}_3,
\end{equation}
with $\phi$ being real. The $\mathcal{R}$ can be thought of as complex conjugation if $\phi$ is vanishing. This gives rise to the orientifold plane at the fixed locus of $\mathcal{R}$. The branes (split into fractional branes) from hidden sector~\cite{Mansha:2025yxm} are utilized in this letter to cancel the untwisted tadpoles.

The (stacks of) D6-branes on this background are described by the wrapping numbers
$(n^i, m^i)$,
the charges under the twisted RR-fields $\epsilon^i\in\{-1,1\}$, the
position $\delta^i\in\{0,1\}$, 
and the discrete
Wilson lines $\lambda^i\in\{0,1\}$.
The D6-brane wraps the one-cycle $n^i [{a'}^i] + m^i [b^i]$ on the $i$-th torus,
the fundamental one-cycles $[{a'}^i]=[a^i]+\frac{\beta^i}{2} [b^i]$ and $[b^i]$ are shown in Fig. \ref{fig:Z2-lattice}.
The $\epsilon^i$ satisfy $\epsilon^1=\epsilon^2 \epsilon^3$, and are related to the discrete Wilson lines~\cite{Sen:1998ii}.
The position is
described by the three parameters $\delta^i$, where $\delta^i=0$ if the D6-brane goes through
fixed point 1 on the $i$-th torus, and $\delta^i=1$ otherwise. An alternative way to characterise a D6-brane is to
use $\epsilon^i_{kl}\in\{-1,0,1\}$, $i\in\{1,2,3\}$, $k,l\in\{1,2,3,4\}$
\cite{Blumenhagen:2005tn}, instead of $\epsilon^i$,
$\delta^i$ and $\lambda^i$. $\epsilon^i_{kl}$ is the charge of the D6-brane under the fixed point labelled $kl$
in the $i$-th twisted sector. The $\epsilon^i_{kl}$ can be determined from $\epsilon^i$,
$\delta^i$ and $\lambda^i$.
Note that for each $i$ only
four out of the sixteen $\epsilon^i_{kl}$ are non-zero.
In both these descriptions there is some redundancy
\cite{Blumenhagen:2005tn}. Rather than fixing some of the $\epsilon^i_{kl}$ charges
to be 1 \cite{Blumenhagen:2005tn}, it will here be more convenient to
choose $n^{1,2} > 0$ (or $m^i$ positive if $n^i$ vanishes).

It is useful to define $\widetilde{m}^i = m^i + \frac{\beta^i}{2} n^i$, such that a D6-brane wraps
the one-cycle $n^i [a^i] + \widetilde{m}^i [b^i]$ on the $i$-th torus.
The volume of this one-cycle
is given by
\begin{eqnarray}
 V^i=\sqrt{(n^i)^2 (R_1^{(i)})^2 + (\widetilde{m}^i)^2 (R_2^{(i)})^2}\,,
 \label{onecyclevolume}
\end{eqnarray}
and the tree-level gauge coupling is
\begin{align}
 \frac{1}{g_\text{tree}^2} &= e^{-\phi_{10}} \prod_i V^i = \frac{e^{-\phi_{4}}}{(T_1T_2T_3)^{1/2}} \prod_i V^i \nonumber\\
 &= \frac{(SU_1U_2U_3)^{1/4}}{(T_1T_2T_3)^{1/2}} \prod_i V^i\, 
 \label{treelevelgc}
\end{align}
where $\phi_{10}$($\phi_{4}$) is the 10(4)-dimensional dilaton in string frame,
$S$ is the dilaton in Einstein frame, $U_i$ are the (real parts of the) complex structure moduli in Einstein frame, and
$T_i = R_1^{(i)} R_2^{(i)}$ are the (real parts of the) K\"ahler moduli. From \eqref{treelevelgc}
one can, using the supersymmetry condition (see below), derive the dependence of the
tree-level gauge kinetic function on the untwisted moduli \cite{Lust:2004cx}
\begin{eqnarray}
 \hat{f}_\text{tree} = S^c n^1 n^2 n^3 - \sum_{i\neq j \neq k=1}^3 U_i^c n^i
 \widetilde{m}^j \widetilde{m}^k \, ,
 \label{treelevelgkf}
\end{eqnarray}
where $S^c$ and $U_i^c$ are the complexified dilaton and complex structure moduli, the axions being RR-fields.
Similarly, $T^c_i$ are complexified K\"ahler moduli with the axions stemming from the NSNS 2-form-field.

The D6-brane is rotated by the angles $\theta^i$, defined via $\tan\theta^i=\widetilde{m}^i R_2^{(i)}/n^i R_1^{(i)}$, with respect to the x-axes of the three tori. Only supersymmetric configurations will be considered in this letter as 
\begin{equation}\label{susy}
\sum_{i=1}^3 \theta^i =0\mod 2\pi  ~.~ \, 
\end{equation}

The charges of the four orientifold planes are denoted as $\eta_{\Omega \mathcal{R}}$ and $\eta_{\Omega \mathcal{R} i}$ with
$i\in \{1,2,3\}$, and have to satisfy
\begin{eqnarray}\label{opsigns}
 \eta_{\Omega \mathcal{R}} \prod_{i=1}^3 \eta_{\Omega \mathcal{R} i} = -1
\end{eqnarray}
in the present case of the $\mathbb{Z}_2\times\mathbb{Z}'_2$ orbifold with $h_{21}=51$ \cite{Blumenhagen:2007ip, Blumenhagen:2005tn}.
The tadpole cancellation conditions are given by
\begin{align}
&\sum_a N_a n_a^1 n_a^2 n_a^3 = 16 \eta_{\Omega \mathcal{R}}, \nonumber\\
&\sum_a N_a n_a^i \widetilde{m}_a^j \widetilde{m}_a^k = -2^{4-\beta^j-\beta^k} \eta_{\Omega \mathcal{R} i},~ i\neq j\neq k \in \{1,2,3\}, \nonumber\\
&\sum_a N_a n_a^i (\epsilon_{a,kl}^i -\eta_{\Omega \mathcal{R}} \eta_{\Omega \mathcal{R} i} \epsilon_{a,\mathcal{R}(k)\mathcal{R}(l)}^i) = 0, \nonumber \\
&\sum_a N_a \widetilde{m}_a^i (\epsilon_{a,kl}^i + \eta_{\Omega \mathcal{R}} \eta_{\Omega \mathcal{R} i} \epsilon_{a,\mathcal{R}(k)\mathcal{R}(l)}^i) = 0 \label{twtadp} ,
\end{align}
where $\mathcal{R}(k)=k$ in case of an untilted torus and $\mathcal{R}(\{ 1,2,3,4\})=\{1,2,4,3\}$ in the other case
\cite{Blumenhagen:2005tn}, and the sum is a sum over all stacks of D6-branes. $N_a$ denotes
the number of D6-branes on stack $a$. The wrapping numbers and twisted charges carry an
index $a$ denoting the D6-brane stack which they describe.
The orientifold projection acts on the wrapping numbers and twisted charges as follows
\begin{eqnarray}
 \widetilde{m}^I &\rightarrow& - \widetilde{m}^I \, ,\nonumber \\
 \epsilon_{kl}^i &\rightarrow& - \eta_{\Omega \mathcal{R}} \eta_{\Omega \mathcal{R} i}
 \epsilon^i_{\mathcal{R}(k)\mathcal{R}(l)}\, .
\end{eqnarray}

\noindent\textbf{A three-family $\mathcal{N}=1$ supersymmetric Pati-Salam models from rigid cycles.} To construct the three family Pati-Salam model using rigid, semi-rigid and non-rigid branes, we follow the strategy outlined in \cite{Forste:2008ex}. In the $\Z_2 \times \Z_2'$ orbifold, the fractional branes invariant under $\Omega \mathcal{R}$ are those placed on top of an exotic O$6^{(+,+)}$ plane that is taken as O$_{\Omega\mathcal{R}}$ in our choice \eqref{opsigns}. The adjoint fields from $aa$ sector do not arise due to the rigid D6-branes.

\begin{table}[!ht]
	\renewcommand{\arraystretch}{1.4}\centering
	$\begin{array}{|c|c|c|}
		\hline
		\text{Stack} & N & (n^1, m^1) \times (n^2, m^2) \times (n^3, m^3) \\
		\hline\hline
		a_1          & 4 & (0, 1)\times (1, 0)\times (0, -1)              \\
		a_2          & 2 & (-4, -1)\times (-2, 1)\times (1, 0)            \\
		a_3          & 2 & (4, 1)\times (2, 1)\times (-3, -1)             \\
		a_4          & 2 & (-1, 0)\times (-2, 1)\times (2, 1)             \\
		a_5          & 2 & (1, 0)\times (0, 1)\times (0, -1)              \\
		\hline
		b_1          & 2 & (1, 0)\times (0, -1)\times (0, 1)              \\
		b_2          & 2 & (-1, 0)\times (0, 1)\times (0, 1)              \\
		\hline
		c_1          & 4 & (0, -1)\times (1, 0)\times (0, 1)              \\
		c_2          & 4 & (0, 1)\times (-1, 0)\times (0, 1)              \\
		\hline
		d_1          & 4 & (0, -1)\times (0, 1)\times (1, 0)              \\
		d_2          & 4 & (0, -1)\times (0, -1)\times (-1, 0)            \\
		\hline
		e_1          & 2 & (1, 0)\times (1, 0)\times (1, 0)               \\
		e_2          & 2 & (-1, 0)\times (-1, 0)\times (1, 0)             \\
		e_3          & 2 & (1, 0)\times (-1, 0)\times (-1, 0)             \\
		e_4          & 2 & (-1, 0)\times (1, 0)\times (-1, 0)             \\
		\hline
	\end{array}$
	\caption{Model with gauge symmetries $\SU(4)_C\times \SU(2)_L\times \SU(2)_1 \times \SU(2)_2\times \SU(2)^3\times \SU(4)^4\times \USp(4)^4$. The torus moduli are $\chi_1=8$, $\chi_2=4$, $\chi_3= 3$, and the tree-level gauge coupling relation is $g_C^2=\frac{5}{4}g_L^2=\frac{25}{44}g_{R}^2=\frac{125}{182} \frac{5}{3} g_Y^2=2 \sqrt{2} \pi e^{\phi_4} $. }
	\label{tab:Model3}
\end{table}

\begin{table}[!ht]
	\renewcommand{\arraystretch}{1.4}\centering
	$\begin{array}{|c|c|c|c|}
		\hline
		\text{Cycles} & \text{$\Theta'$-sector} & \text{$\Theta\Theta'$-sector} & \text{$\Theta$-sector} \\
		\hline
		\Pi_{a_1}     & ((1,3), (1,2))          & ((1,2), (1,2))                & ((1,2), (1,3))         \\
		\hline
		\Pi_{a_2}     & ((1,2), (1,3))          & ((1,2), (1,3))                & ((1,2), (1,2))         \\
		\hline
		\Pi_{a_3}     & ((1,2), (1,2))          & ((1,3), (1,2))                & ((1,3), (1,2))         \\
		\hline
		\Pi_{a_4}     & ((1,2), (1,4))          & ((1,2), (1,4))                & ((1,2), (1,2))         \\
		\hline
		\Pi_{a_5}^*   & ((3,4), (3,4))          & ((2,4), (3,4))                & ((2,4), (3,4))         \\
		\hline
		\Pi_{b_1}^*   & ((3,4), (3,4))          & ((2,4), (3,4))                & ((2,4), (3,4))         \\
		\Pi_{b_2}^*   & ((3,4), (3,4))          & ((2,4), (3,4))                & ((2,4), (3,4))         \\
		\hline
		\Pi_{c_1}^*   & ((2,4), (3,4))          & ((3,4), (3,4))                & ((3,4), (2,4))         \\
		\Pi_{c_2}^*   & ((2,4), (3,4))          & ((3,4), (3,4))                & ((3,4), (2,4))         \\
		\hline
		\Pi_{d_1}^*   & ((3,4), (2,4))          & ((3,4), (2,4))                & ((3,4), (3,4))         \\
		\Pi_{d_2}^*   & ((3,4), (2,4))          & ((3,4), (2,4))                & ((3,4), (3,4))         \\
		\hline
		\Pi_{e_1}^*   & ((2,4), (2,4))          & ((2,4), (2,4))                & ((2,4), (2,4))         \\
		\Pi_{e_2}^*   & ((2,4), (2,4))          & ((2,4), (2,4))                & ((2,4), (2,4))         \\
		\Pi_{e_3}^*   & ((2,4), (2,4))          & ((2,4), (2,4))                & ((2,4), (2,4))         \\
		\Pi_{e_4}^*   & ((2,4), (2,4))          & ((2,4), (2,4))                & ((2,4), (2,4))         \\
		\hline
	\end{array}$ 
	\caption{The hidden sector D6-branes, shown with asterisks are deformed to move away from the origin in order to remove the associated exotic particles.}
	\label{model3fp}
\end{table}

\begin{table}[!ht]
	\centering\renewcommand{\arraystretch}{1.2}\centering
	$\begin{array}{|c|c|c|}
		\hline
		\text{Sector} & \multicolumn{2}{c|}{\SU(4)\times \SU(2)_L\times \SU(2)_1 \times \SU(2)_2\times \text{Hidden~~~~}} \\
		\hline
		a_1 a_2                      & 3\times \left(4,\bar{2},1,1,1,1,1,1,1,1,1,1,1,1,1\right)              & F^i_L (Q_L, L_L)            \\
		a_1 a_3'                     & 6\times \left(\bar{4},1,\bar{2},1,1,1,1,1,1,1,1,1,1,1,1\right)        & F^i_R (Q_R, L_R)            \\
		a_1 a_3                      & 1\times \left(4,1,\bar{2},1,1,1,1,1,1,1,1,1,1,1,1\right)              & F^{c}_R (Q_R, L_R)          \\
		a_1 a_4                      & 2\times\left(4,1,1,\bar{2},1,1,1,1,1,1,1,1,1,1,1\right)               & F^{c\prime i}_R (Q_R, L_R) \\
		a_{2_{\overline{\yng(1,1)}}} & 4\times (1,\bar{1}_{\overline{\yng(1,1)}},1,1,1,1,1,1,1,1,1,1,1,1,1)  & S_L^i                       \\
		a_2 a_3                      & 3\times \left(1,\bar{2},2,1,1,1,1,1,1,1,1,1,1,1,1\right)              & \Phi'_i(H'_u, H'_d)         \\
		a_2 a_3'                     & 5\times \left(1,2,2,1,1,1,1,1,1,1,1,1,1,1,1\right)                    & \Phi_i(H_u^i, H_d^i)        \\
		a_2 a_4'                     & 1\times \left(1,2,1,2,1,1,1,1,1,1,1,1,1,1,1\right)                    & \Xi_i                       \\
		a_{3_{\overline{\yng(1,1)}}} & 60\times (1,1,\bar{1}_{\overline{\yng(1,1)}},1,1,1,1,1,1,1,1,1,1,1,1) & S_R^i                       \\
		a_{3_{\overline{\yng(2)}}}   & 6\times (1,1,\bar{3}_{\overline{\yng(2)}},1,1,1,1,1,1,1,1,1,1,1,1)    & T_R^i                       \\
		a_3 a_4'                     & 3\times \left(1,1,2,2,1,1,1,1,1,1,1,1,1,1,1\right)                    & \Delta_i                    \\  
		a_2a_5'                      & 1\times (1,2,1,1,2,1,1,1,1,1,1,1,1,1,1)                               & X_L^{i}                     \\
		a_2b_1                       & 1\times (1,2,1,1,1,\bar{2},1,1,1,1,1,1,1,1,1)                         & X_L^{i}                     \\
		a_2b_2'                      & 1\times (1,2,1,1,1,1,2,1,1,1,1,1,1,1,1)                               & X_L^{i}                     \\
		a_2c_1'                      & 2\times (1,\bar{2},1,1,1,1,1,\bar{4},1,1,1,1,1,1,1)                   & X_L^{i}                     \\
		a_2c_2                       & 2\times (1,\bar{2},1,1,1,1,1,1,4,1,1,1,1,1,1)                         & X_L^{i}                     \\
		a_3a_5'                      & 2\times (1,1,\bar{2},1,\bar{2},1,1,1,1,1,1,1,1,1,1)                   & X_R^{i}                     \\
		a_3a_5                       & 1\times (1,1,\bar{2},1,2,1,1,1,1,1,1,1,1,1,1)                         & X_R^{i}                     \\
		a_3b_1'                      & 1\times (1,1,\bar{2},1,1,\bar{2},1,1,1,1,1,1,1,1,1)                   & X_R^{i}                     \\
		a_3b_1                       & 2\times (1,1,\bar{2},1,1,2,1,1,1,1,1,1,1,1,1)                         & X_R^{i}                     \\
		a_3b_2'                      & 2\times (1,1,\bar{2},1,1,1,\bar{2},1,1,1,1,1,1,1,1)                   & X_R^{i}                     \\
		a_3b_2                       & 1\times (1,1,\bar{2},1,1,1,2,1,1,1,1,1,1,1,1)                         & X_R^{i}                     \\
		a_3c_1'                      & 2\times (1,1,\bar{2},1,1,1,1,\bar{4},1,1,1,1,1,1,1)                   & X_R^{i}                     \\
		a_3c_1                       & 4\times (1,1,\bar{2},1,1,1,1,4,1,1,1,1,1,1,1)                         & X_R^{i}                     \\
		a_3c_2'                      & 4\times (1,1,\bar{2},1,1,1,1,1,\bar{4},1,1,1,1,1,1)                   & X_R^{i}                     \\
		a_3c_2                       & 2\times (1,1,\bar{2},1,1,1,1,1,4,1,1,1,1,1,1)                         & X_R^{i}                     \\
		a_4c_1'                      & 1\times (1,1,1,\bar{2},1,1,1,\bar{4},1,1,1,1,1,1,1)                   & X_R^{i}                     \\
		a_4c_2'                      & 1\times (1,1,1,\bar{2},1,1,1,1,\bar{4},1,1,1,1,1,1)                   & X_R^{i}                     \\
		a_3d_1'                      & 2\times (1,1,\bar{2},1,1,1,1,1,1,\bar{4},1,1,1,1,1)                   & X_R^{i}                     \\
		a_3d_1                       & 2\times (1,1,\bar{2},1,1,1,1,1,1,4,1,1,1,1,1)                         & X_R^{i}                     \\
		a_3d_2'                      & 2\times (1,1,\bar{2},1,1,1,1,1,1,1,\bar{4},1,1,1,1)                   & X_R^{i}                     \\
		a_3d_2                       & 2\times (1,1,\bar{2},1,1,1,1,1,1,1,4,1,1,1,1)                         & X_R^{i}                     \\
		a_4d_1'                      & 1\times (1,1,1,2,1,1,1,1,1,4,1,1,1,1,1)                               & X_R^{i}                     \\
		a_4d_2                       & 1\times (1,1,1,2,1,1,1,1,1,1,\bar{4},1,1,1,1)                         & X_R^{i}                     \\
		a_3e_2'                      & 1\times (1,1,2,1,1,1,1,1,1,1,1,1,4,1,1)                               & X_R^{i}                     \\
		a_3e_2                       & 1\times (1,1,2,1,1,1,1,1,1,1,1,1,\bar{4},1,1)                         & X_R^{i}                     \\
		\hline
	\end{array}$
	\caption{The chiral particle spectrum for the model in Table~\ref{tab:Model3}, and the hidden part of the symmetry is $\SU(2)^3\times \SU(4)^4\times \USp(4)^4 $. The chiral spectrum is singlet under this symmetry.}
	\label{tab:spectrum3}
\end{table}

We present the model in Table~\ref{tab:Model3} with the corresponding fixed points of the brane stacks given in Table~\ref{model3fp}. We deform the hidden sector D6-branes to move away from the origin in order to get rid of some of the associated exotic particles. The complete perturbative particle spectrum is shown in Table~\ref{tab:spectrum3}. The several exotic particles $X_L^{i}$ and $X_R^{i}$ can be decoupled from low energy spectrum via strong-dynamics as discussed in the companion paper~\cite{Mansha:2025yxm}.
 
The additional constraint from (\ref{twtadp}) makes it very difficult to construct the consistent vacua. Indeed, the model reported in~\ref{tab:Model3} is of phenomenological interest due to the breaking of the Pati-Salam gauge symmetry down to the SM gauge symmetry. The purpose of the present section is to give an example of these more involved constructions which also admit the low energy theories very closer to realistic particle physics. Models presented in~\cite{Mansha:2025yxm} usually do not have the Higgs fields to break the Pati-Salam gauge symmetry down to the SM. We address this important feature of model building for the Model in Table~\ref{tab:Model3}. With the choice of the fixed points listed in Table~\ref{model3fp}, one can calculate the chiral spectrum presented in Table~\ref{tab:spectrum3}. The Table~\ref{tab:Model3} consists of five sets D6-branes, i.e., ${a, b, c, d, e}$. The set $a$ consists of five stacks of fractional D6-branes, but each with different bulk wrapping numbers. The set $a$, most importantly, gives rise to the gauge symmetries and chiral spectrum. In contrast to the set $a$, the sets $b$, $c$, $d$, and $e$ are the bulk the D6-branes which are split into fractional constituents. It can be checked that The table~\ref{tab:Model3} satisfies all the constraints presented in~(\ref{twtadp}) for untwisted and twisted tadpoles, and~(\ref{susy}) for supersymmetry condition. 

The gauge symmetries derived from the set $a$ are $\SU(4)\times \SU(2)_L\times \SU(2)_1 \times \SU(2)_2\times \SU(2)$, which give rise to the Pati-Salam-like theory from the subsets $a_1$, $a_2$, $a_3$, and $a_4$.
In contrast to set $a$, the sets $b$, $c$, $d$, and $e$ yield the gauge symmetries $\U(2)^2$, $\U(4)^2$, $\U(4)^2$ and $\USp(4)^4$ respectively. The latter gauges groups from the fractional D6-branes can be deformed into bulk D6-branes upon Higgsings as follows
\begin{equation}
    \{b_1, b_2\}, \{c_1, c_2\}, \{d_1, d_2\}, \{e_1, e_2, e_3, e_4\} \to b, c, d, e
\end{equation}
\begin{align}\label{higgsingm3}
&\U(2)^2\times \U(4)^2\times \U(4)^2\times \USp(4)^4 \nonumber\\
&\qquad \qquad \longrightarrow \; \U(1)\times \U(2)\times \U(2)\times \USp(4).
\end{align}
Some of the $\U(1)$s are not really gauge symmetries, but only global ones because their gauge bosons would receive Stueckelberg mass through the Green-Schwarz mechanism~\cite{Cvetic:2001nr}. Thus, Pati-Salam gauge symmetries $\SU(4)\times \SU(2)\times \SU(2)$ are obtained from D-branes $a_1$, $a_2$, $a_3$, and $a_4$, while the extra symmetries $\U(2)^2\times \U(4)^2\times \U(4)^2\times \USp(4)^4$ arise from the set $b$, $c$, $d$, and $e$ which turns out to be, after Higgsing~(\ref{higgsingm3}), the gauge symmetries $\U(1)\times \U(2)\times \U(2)\times \USp(4)$. 

The rigid brane $a_4$ is moved into visible sector for the sake of symmetry breaking, thus modifying the condition for three generations,
\begin{equation}\label{3family1}
    I_{a_1a_2} + I_{a_1a_2'} = \pm3,
\end{equation}
\begin{equation}\label{3family2}
    I_{a_1a_3} + I_{a_1a_3'} + I_{a_1a_4} + I_{a_1a_4'} = \mp3.
\end{equation}
We present the chiral spectrum of this theory in the Table~\ref{tab:spectrum3}. In particular, we present the conventions of some particles concretely, for example, the SM fermions, and Higgs fields, etc. Moreover, the visible sector will suffer from the uncancelled twisted tadpoles. Therefore, an additional D6-brane $a_5$ is added to cancel the latter. In order to get the consistent model, untwisted tadpoles are canceled by adding more semi-rigid branes from the set $b$, $c$, $d$, and $e$ such that they do not make any contribution to the twisted tadpoles. \\

\noindent\textbf{Phenomenological Study.} 
The gauge symmetries from $a_1$, $a_2$, $a_3$, and $a_4$ are $\U(4)_C$, $\U(2)_L$, $\U(2)_1$, and $\U(2)_2$, respectively. The anomalies from four global $\U(1)$s of 
$\U(4)_C$, $\U(2)_L$, $\U(2)_1$, and $\U(2)_2$ are cancelled by the Green-Schwarz mechanism, and the gauge fields of 
these $\U(1)$s obtain masses via the linear $B\wedge F$ couplings. So, the
effective gauge symmetry is $\SU(4)_C\times \SU(2)_L\times \SU(2)_1 \times \SU(2)_2$. When $\Delta_i$ obtain Vacuum Expectation Values (VEVs), the $\SU(2)_1 \times \SU(2)_2$ gauge symmetry is broken down to the diagonal $\SU(2)_R$ gauge symmetry. Thus, we obtain the Pati-Salam  $\SU(4)_C\times \SU(2)_L\times \SU(2)_R$ gauge symmetry. We have three families of left-handed fermions $F^i$, and three families of right-handed fermions from $F_R^i$, $F_R^c$, $F_R^{c\prime i}$, i.e., $6-1-2=3$.

Moreover, by giving VEVs to $F_R^c$ and a linear combination of $F_R^i$,
we can break the Pati-Salam gauge symmetry down to the SM gauge symmetry and preserve the four-dimensional $\mathcal{N}=1$ supersymmetry by keeping the D-flatness and F-flatness. In addition, we can generate the SM fermion masses and mixing,
as well as the vector-like particle masses for $F_R^{\prime ci}$ and two other linear combinations of $F_R^i$ via the following superpotential
\begin{equation}
    W \supset y_{ij}^k F_L^i F_R^j \Phi_k + y_{ij}^{\prime k} F_R^{c\prime i}F_R^j \Delta_k \, .
\end{equation}
Furthermore, the $\SU(4)$ gauge symmetries in the hidden sector have negative beta functions, and then the supersymmetry can be broken via gaugino condensations. 

In intersecting D-brane models, proton decay mediating operators are absent at all orders in perturbation theory due to worldsheet consistency conditions. All interactions originate from open string diagrams, which require an even number of string endpoints and conservation of brane charges. As a result, processes involving an odd number of SU(3) triplets, such as $U + D + D \rightarrow \text{leptons}$, cannot be realized perturbatively~\cite{Aldazabal:2000dg}. To be concrete, the gauge symmetry of our Pati-Salam models is $U(4)_C\times U(2)_L \times U(2)_R$, and thus we have three anomalous $U(1)$ gauge symmetries $U(1)_C \times U(1)_L \times U(1)_R$ whose gauge anomalies are cancelled by the generalized Green-Schwarz mechanism. The proton decay operators such as $QQQL$ and $U^c D^c U^c E^c$ are forbidden by these anomalous $U(1)$ gauge symmetries.

The tree-level gauge coupling relations for the model in Table~\ref{tab:Model3} are
\begin{equation}
g_C^2 = \frac{5}{4} g_L^2 = \frac{25}{44} g_R^2 = \frac{125}{182} \cdot \frac{5}{3} g_Y^2,
\end{equation}
where the weak hypercharge arises from a canonical Pati-Salam embedding involving the $\SU(4)_C$ and $\SU(2)_{1,2}$ generators. This implies
\begin{equation}
g_L^2 = \frac{250}{273} g_Y^2 ~~ \Rightarrow ~~ \sin^2 \theta_W = \frac{g_Y^2}{g_Y^2 + g_L^2} = \frac{273}{523} \approx 0.521, \nonumber
\end{equation}
valid at the string/unification scale. While this differs from the observed low-energy value $\sin^2 \theta_W(M_Z) \approx 0.231$, 
the renormalization group equation running by considering the vector-like particles and adjoint fields~\cite{Li:2022cqk} as well as string threshold corrections~\cite{Lust:2003ky, Ghilencea:2000dg} can bridge the gap. In fact, the hypercharge normalization $k_Y$, defined via $g_Y^2 = k_Y g^2$, is not fixed universally in D-brane models, as $U(1)_Y$ typically arises from a linear combination of $U(1)_a$ factors with model-dependent coefficients and wrapping numbers~\cite{Ibanez:2001nd, Blumenhagen:2005mu}. By choosing alternative, anomaly-free hypercharge embeddings, it is possible to adjust $k_Y$ to bring the tree-level prediction of $\sin^2 \theta_W$ closer to the MSSM or SU(5) GUT value $\sin^2 \theta_W(M_{\text{GUT}}) = 3/8$. Additionally, kinetic mixing and generalized Chern-Simons terms can lead to effective shifts in the gauge couplings and $k_Y$ at low energies~\cite{Anastasopoulos:2006cz}. 

Finally, the string scale $M_\text{S}$, which is determined by the dilaton and internal volumes, can be tuned to stay in the range of $M_\text{S} \sim 10^{14} - 10^{17}\,$GeV to remain compatible with phenomenological constraints on gauge coupling unification and electroweak observables.  \\

\noindent\textbf{Conclusions.}
We have presented a three-generation $\mathcal{N}=1$ supersymmetric Pati-Salam models from rigid intersecting D6-branes on $\T^6/(\Z_2\times \Z_2')$ orientifold with discrete torsion. The model is first of its kind constructed in the factorizable compactification lattice with moduli stabilization. The construction is particularly simple as it only involves rectangular two-tori. 

We showed that the Pati-Salam gauge symmetry can be broken down to the SM gauge symmetry via the supersymmetry preserving Higgs mechanism, the SM fermion masses and mixings can be generated, and the supersymmetry can be broken via gaugino condensations in the hidden sector. 
We have presented the complete perturbative particle spectrum, tree-level gauge coupling relations. Further detailed phenomenology of the model needs a thorough analysis of the supersymmetry breaking soft terms, and the allowed masses of the SM fermions with moduli stabilization. \\

\noindent\textbf{Acknowledgments.}
We thank the anonymous referee for their valuable suggestions, which have significantly improved the presentation of the paper. AM is supported by the Guangdong Basic and Applied Basic Research Foundation (Grant No. 2021B1515130007), Shenzhen Natural Science Fund (the Stable Support Plan Program 20220810130956001). TL is supported in part by the National Key Research and Development Program of China Grant No. 2020YFC2201504, by the Projects No. 11875062, No. 11947302, No. 12047503, and No. 12275333 supported by the National Natural Science Foundation of China, by the Key Research Program of the Chinese Academy of Sciences, Grant No. XDPB15, by the Scientific Instrument Developing Project of the Chinese Academy of Sciences, Grant No. YJKYYQ20190049, and by the International Partnership Program of Chinese Academy of Sciences for Grand Challenges, Grant No. 112311KYSB20210012. MS is supported in part by the National Natural Science Foundation of China (Grant No. 12475105).

 
%

\end{document}